\documentclass[pdflatex,sn-mathphys-num]{sn-jnl}
\usepackage{graphicx}%
\usepackage{multirow}%
\usepackage{amsmath,amssymb,amsfonts}%
\usepackage{physics}
\usepackage{amsthm}%
\usepackage{mathrsfs}%
\usepackage[title]{appendix}%
\usepackage{xcolor}%
\usepackage{textcomp}%
\usepackage{manyfoot}%
\usepackage{booktabs}%
\usepackage{algorithm}%
\usepackage{algorithmicx}%
\usepackage{algpseudocode}%
\usepackage{listings}%
\usepackage{natbib}%
\usepackage{caption}

\theoremstyle{thmstyleone}%

\theoremstyle{thmstyletwo}%

\theoremstyle{thmstylethree}%

\begin{document}



\title[Article Title]{Fast leave-one-cluster-out cross-validation using clustered Network Information Criterion (NICc)} 

\author*[1,2]{\fnm{Jiaxing}\sur{Qiu}}\email{jq2uw@virginia.edu}

\author[2]{\fnm{Douglas E.}\sur{Lake}}\email{del2k@uvahealth.org}

\author[2]{\fnm{Pavel}\sur{Chernyavskiy}}\email{pchern@virginia.edu}

\author[1,3]{\fnm{Teague R.}\sur{Henry}}\email{ycp6wm@virginia.edu}

\affil[1]{\orgdiv{School of Data Science}, \orgname{University of Virginia}}

\affil[2]{\orgdiv{School of Medicine}, \orgname{University of Virginia}}


\affil[3]{\orgdiv{Department of Psychology}, \orgname{University of Virginia}}

\abstract{ 

For prediction models developed on clustered data that do not account for cluster heterogeneity in model parameterization, it is crucial to use cluster-based validation to assess model generalizability on unseen clusters. This paper introduces a clustered estimator of the Network Information Criterion (NICc) to approximate leave-one-cluster-out deviance for standard prediction models with twice-differentiable log-likelihood functions. The NICc serves as a fast alternative to cluster-based cross-validation. 

Stone (1977) proved that the Akaike Information Criterion (AIC) is asymptotically equivalent to leave-one-observation-out cross-validation for true parametric models with independent and identically distributed (i.i.d.) observations. Ripley (1996) noted that the Network Information Criterion (NIC), derived from Stone's proof, is a better approximation when the model is misspecified. For clustered data, we derived NICc by substituting the Fisher information matrix in the NIC with a clustering-adjusted estimator. The NICc imposes a greater penalty when the data exhibits stronger clustering, thereby allowing the NICc to better prevent over-parameterization.

In a simulation study and an empirical example, we used standard regression to develop prediction models for clustered data with Gaussian or binomial responses. Compared to the commonly used AIC and BIC for standard regression, NICc provides a much more accurate approximation to leave-one-cluster-out deviance and results in more accurate model size and variable selection, as determined by cluster-based cross-validation, especially when the data exhibit strong clustering.


}


\maketitle

\section{Introduction} \label{sec_intro}

Our article concerns the development of predictive models in a data-rich biomedical settings. 
These models are typically trained on repeatedly measured patient medical records 
to predict various patient outcomes, such as mortality, adverse medical events (e.g., sepsis), re-hospitalization, and length of stay. Such biomedical data are often grouped into clusters (typically, patients) where observations are assumed to be correlated within each cluster but independent across clusters \cite{rogers_regression_1994}. The repeated measures, such as laboratory and monitoring data from each patient, are usually correlated within each patient but, in the absence of a second grouping factor (e.g., facility), are assumed to be independent across patients \cite{harrell_overview_2015}. Another example considers a medical facility as a clustering unit. In multi-center studies, patients treated in the same facility may be more dependent due to consistent treatment policies and similar patient populations, whereas patients from different centers may differ due to varying policies \cite{bouwmeester_prediction_2013}.

When developing prediction models on such clustered data, the primary objective is to optimize the predictive performance of a model on unseen clusters, such as a new or external patient population  \citep{debray_transparent_2023}, rather than to perform inference on the estimated parameters. In this article, to mirror the clinical quantities of interest, we focus on population-level prediction rather than cluster-specific inference and prediction. We term model predictive performance on unseen clusters as \textbf{out-of-cluster performance}. 
In the context of population-level prediction of outcomes in the yet unseen clusters, commonly employed standard regression models, such as traditional generalized linear models (GLMs), and more complex machine learning or deep learning techniques neglect the within-cluster correlation in the model parameterization and focus instead on achieving high out-of-cluster predictive performance \citep{bouwmeester_reporting_2012, reddy_healthcare_2015, takada_internal-external_2021}.

The motivation of this article stems from the need for an efficient model selection criterion that can accurately approximate out-of-cluster predictive performance when using standard regression applied to clustered data. In practice, the datasets are often massive ($>1$ million rows) and contain high-dimensional feature spaces, as we present in more detail in subsection \ref{subsec_1}. This usually makes complex methods that explicitly model cluster heterogeneity and correlation computationally infeasible. Likewise, to perform feature selection with cross-validation, a model would need to be estimated many times on the same massive dataset, further highlighting the need for increased computational efficiency. For this, we propose the \textit{clustered} Network Information Criterion (NICc) to more efficiently develop prediction models on clustered data.

The rest of the article is organized as follows. We dedicate the remainder of the introduction (Section \ref{sec_intro}) to detailing the usage of standard regression in clinical predictive modeling compared to other methods (Section \ref{sec_intro_part1}), and the limitations of using traditional information criteria to approximate the out-of-cluster predictive performance of such models (Section \ref{sec_intro_part2}). Then we introduce the \textit{clustered} Network Information Criterion (NICc) as a fast leave-one-cluster-out cross-validation estimator in section \ref{sec_nicc}. Section \ref{sec_simu} and \ref{sec_simu_res} compare our proposed criterion to other common model selection criteria via a Monte Carlo simulation. Section \ref{sec_empirical} presents an application of our criterion to daily observations on neonates to predict mortality. Finally, Section \ref{sec_discuss} contains a discussion of our new method. 

\subsection{Predictive models for clustered biomedical data} \label{sec_intro_part1}

\subsubsection{Usage of standard regression} \label{subsec_1}

In practice, GLMs that overlook the correlation in the data are widely used for prediction (either diagnostic or prognostic) on clustered medical data \cite{shipe_developing_2019, steyerberg2008clinical}. This approach is preferred for various reasons, including non-inferior predictive performance compared to complex machine learning techniques \cite{christodoulou_systematic_2019, eftekhar_comparison_2005}, ease of interpretation \cite{boateng_review_2019}, computational efficiency with large datasets \cite{niestroy_discovery_2022, qiu_highly_2024}, and feasibility in real-time monitoring implementations \cite{fairchild_hero_2012, ruminski_impact_2019}.

Logistic regression models developed using clustered patient data have served as the basis for many diagnostic or prognostic prediction models that have been adopted for clinical use \cite{shipe_developing_2019}. For example, in lung cancer prediction, the Tammemagi model \cite{mcwilliams_probability_2013, ostrowski_performance_2021} used a logistic regression model developed on repeatedly measured pulmonary nodules in patient computed tomography scan. To predict mortality in neonatal intensive care units (NICUs), the heart rate characteristics (HeRO) score \cite{fairchild_hero_2012, ibrahim_predictive_2023, bennaoui_hero_2024} implemented in multiple bedside monitoring systems was based on a logistic regression, using heart rate characteristics of the electrocardiogram. In sepsis and bloodstream infection prediction, penalized logistic regression models \cite{zimmet_pathophysiologic_2020, qiu_pathophysiological_2023, kausch_cardiorespiratory_2023} have leveraged patient demographics, repeated laboratory results, and vital sign data, effectively predicting both recurrent infections for observed patients and new events for out-of-sample patients. 

When the final prediction model needs to be selected from numerous candidates in a high-dimensional predictor space, standard regression methods are computationally efficient, particularly when cross-validation or bootstrap methods are used for model assessment. For example, a highly comparative time series analysis with over 5000 features \cite{fulcher_highly_2014} was applied to $>18$ million 10-minute heart rate and oxygen saturation segments from $>6000$ infants to predict mortality \cite{niestroy_discovery_2022}, and $>7$ million 10-minute vital sign segments from $>700$ preterm infants to predict adverse cardiorespiratory outcomes \cite{qiu_highly_2024}. Logistic regression with cluster-based cross-validation significantly expedited the identification of top predictors for these outcomes. Training cluster-heterogeneity-adjusted random effect (RE) models on such high-dimensional massive datasets with cross-validation would be computationally inefficient and likely to encounter convergence issues. Furthermore, standard regression models have been successfully implemented in various real-time cardiorespiratory monitoring systems, such as the HeRO score for monitoring NICU sepsis and mortality \cite{fairchild_hero_2012} and the CoMET score for detecting clinical deterioration in hospitalized patients \citep{ruminski_impact_2019, monfredi_novel_2023}.

\subsubsection{Usage of random effect and Generalized Estimating Equations methods} \label{subsec_2}

RE methods \cite{skrondal_generalized_2004} and generalized estimating equation (GEE) methods \cite{liang_longitudinal_1986} are recommended for addressing between-cluster heterogeneity, modeling within-cluster correlation structures, and making valid inferences on estimated model parameters \cite{gardiner_fixed_2009, steyerberg2008clinical}. 
However, in the context of population-level prediction of biomedical outcomes in the out-of-sample clusters, such methods are less common in practice because subject-specific inference is not needed \cite{bouwmeester_reporting_2012, bouwmeester_prediction_2013}.

Besides the known limitations of RE models in practice, including 
difficulty in specifying a complete likelihood function for real world problems \citep{zorn_generalized_2001}, and various convergence issues \citep{van_der_elst_unbalanced_2016, mcneish_covariance_2020, nie_convergence_2007}, a comparative study by \citet{bouwmeester_prediction_2013} showed that logistic regression models developed on clustered data demonstrate non-inferior out-of-cluster performance vs. random intercept models. Otherwise, few studies have explicitly compared the out-of-cluster performance of RE and GEE methods vs. standard GLMs, also considering the additional computing time required for cluster-specific estimation and prediction. We conducted this comparison as a secondary objective in this study.

\subsection{Evaluating out-of-cluster predictive performance} \label{sec_intro_part2}

Cross-validation is often considered the ``standard" procedure to evaluate the predictive accuracy of a model \cite{hastie_model_2009, ripley_pattern_1996}. 
When assessing the out-of-\textit{cluster} performance of a model, leave-one-\textit{observation}-out cross-validation \cite{stone_cross-validatory_1974, browne_cross-validation_2000} will result in a biased (overly optimistic) estimation due to the dependence between the left-out observation and the observations from the same cluster \cite{rabinowicz_cross-validation_2022}. Model validation procedures for clustered data, such as cross-validation and bootstrapping, are based on resampling the clusters instead of observations to preserve between-cluster heterogeneity during assessment \cite{feng_comparison_1996, bergmeir_use_2012}. The leave-one-\textit{cluster}-out cross-validation, also known as internal-external cross-validation, is commonly used for clustered data \cite{takada_internal-external_2021, de_jong_developing_2021}. This procedure iteratively uses all but one cluster for model development and assesses performance in the remaining cluster. 

When the number of clusters is extensive, leave-one-cluster-out cross-validation becomes computationally inefficient and/or infeasible. Thus, an information criterion approximation can be a fast alternative. The Akaike Information Criterion (AIC) \cite{akaike_information_1998} and Bayesian Information Criterion (BIC) \cite{schwarz_estimating_1978} are the two most commonly used frequentist criteria for regression models estimated via maximizing the log-likelihood (MLE) \cite{calcagno2010glmulti, brewer_relative_2016}. AIC aims to approximate the expected Kullback-Leibler information between the true model and a selected model whose parameters are estimated by maximum likelihood \cite{burnham_multimodel_2004}. It penalizes the log likelihood by the number of estimated parameters. BIC aims to select the candidate model which is a posterior most probable \cite{neath_bayesian_2012}. It penalizes the log likelihood by the number of estimated parameters multiplied by the log of the sample size. Extensions of AIC and BIC have been developed for models that account for cluster heterogeneity in model parameterization, particularly in RE  \cite{vaida_conditional_2005, jones_bayesian_2011} and GEE \cite{pan_akaikes_2001, wang_consistent_2009}. 

For selecting a model for prediction, AIC has a desirable property of approximating leave-one-observation-out cross-validation for i.i.d. samples \cite{stone_asymptotic_1977}, but its accuracy in approximating cluster-based cross-validation is expected to be worse since the i.i.d assumption is violated. On the other hand, BIC imposes a greater penalty than AIC by incorporating sample size into the penalty term, often resulting in oversimplified and/or misspecified models for larger sample sizes \cite{jones_bayesian_2011}. This could be more problematic for clustered data. When the total number of clusters is fixed but more observations are collected over a longer period of time, such as vital sign data continuously measured for the same NICU population, this larger penalty in BIC may over-penalize reasonable candidate models simply due to more observations per cluster. Therefore, it is important to investigate a new information criterion that can efficiently assess out-of-cluster performance and overcome the limitations of these traditional information criteria, in the context of clustered biomedical data.

\section{The clustered Network Information Criterion} \label{sec_nicc}

Here, we introduce the clustered Network Information Criterion (NICc) as a fast alternative to leave-one-cluster-out cross-validation for standard prediction models with twice-differentiable log-likelihood functions. 

\citet{stone_asymptotic_1977} proved that AIC is asymptotically equivalent to leave-one-observation-out cross-validation if the parametric model is correctly specified and the sample consists of independent and identically distributed (i.i.d) observations. \citet{ripley_pattern_1996} pointed out that the Network Information Criterion (NIC) \cite{murata_network_1994}, derived from Stone's proof, is a better approximation when the model is misspecified.
The essential results to illustrate the asymptotic equivalence between NIC and leave-one-out cross-validated deviance are listed in the next subsection, but detailed proof can be found in \cite{stone_asymptotic_1977} and \cite[p.~71]{ripley_pattern_1996}. We extend the standard NIC to a clustered estimator of the Network Information Criterion (that we term the NICc). The NICc is designed to approximate the leave-one-cluster-out deviance (looDeviance) for standard regression models with twice-differentiable log-likelihood functions, that is to say, positive semi-definite Hessian matrices. Inspired by the cluster adjustment used in the Huber sandwich estimator to estimate variance-covariance matrix for clustered data \cite{harrell_overview_2015, freedman_so-called_2006, barlow_robust_1994, mccullagh_generalized_2019}, we estimate the NICc by substituting the Fisher information matrix in the previously-derived NIC with its estimator that adjusts for clustering. Derivations are provided as follows.

\subsection{Network Information Criterion}

Adopting the notation in \citet{stone_asymptotic_1977} and \citet{ripley_pattern_1996}, let the data be denoted as $\{(X_i, y_i), i=1,...,N\} $, where $X_i$ is the $i^{\text{th}}$ vector of predictor variables, $y_i$ is the $i^{\text{th}}$ univariate response. $p(y_{i}| X_i, \Theta)$ is the likelihood function of $y_{i}$ given $X_i$ and the parameter vector $\Theta$ of length $p$. 

The log-likelihood assessment \cite{stone_asymptotic_1977} of all $N$ observations is:
\[
L(\hat{\Theta}) = \sum_{i}\log p(y_i | X_i, \hat{\Theta}),
\] where $\hat{\Theta}$ is the MLE estimate of $\Theta$ using all observations, $\log p(y_i | X_i, \hat{\Theta})$ is the predicted log-likelihood of the $i^{\text{th}}$ observation under $\hat{\Theta}$.

The leave-one-observation-out log-likelihood assessment, denoted by $A$, is:
\begin{equation*}
  A = \sum \log p(y_{-i} | X_{-i}, \Tilde{\Theta}),
\end{equation*}
where $\Tilde{\Theta}$ is MLE estimate of $\Theta$ using all observations except for the $i^{\text{th}}$ observation, $\log p(y_{-i} | X_{-i}, \Tilde{\Theta})$ is the predicted log-likelihood of individual observations except for the $i^{\text{th}}$ under $\Tilde{\Theta}$. 

Suppose $\log p(y|X,\Theta)$ is twice-differentiable with respect to $\Theta$, let $J$ denote the $p\times p$ Hessian matrix of the log-likelihood function, and let $K$ denote the Fisher information matrix:
\begin{equation*}
    J = -E_{p} (\pdv{\log p(y|X,\Theta_0) }{\theta}{\theta^{T}})
    \; \; \text{and} \; \;
    K = \mathrm{cov}_{p} (\pdv{\log p(y|X,\Theta_0)}{\theta}).
\end{equation*}
\citet{stone_asymptotic_1977} proved that asymptotically ($N \to \infty$), 
\begin{equation} \label{eq_loo}
    A = L(\hat{\Theta}) - \text{trace}[J^{-1}K].
\end{equation} 
Equation (\ref{eq_loo}), multiplied by $-2$, is the Network Information Criterion (NIC) \cite{murata_network_1994, ripley_pattern_1996}:
\begin{equation} \label{eq_NIC}
\text{NIC} = -2L(\hat{\Theta}) + 2\text{trace}[J^{-1}K].
\end{equation} 
When sample size N is finite but essentially large, NIC can be estimated by the sample estimators of Hessian matrix $J$ and Fisher information matrix $K$ \cite[p.~71]{ripley_pattern_1996}.  

Note that if the parametric family contains the true model, $J = K$\cite{cox_theoretical_1979}. Then, $A = L(\hat{\Theta}) - p$ because $\text{trace}[J^{-1}K] = \text{trace}[I_{p \times p}]$. Multiplying by $-2$, we have the Akaike Information Criterion: 
$\text{AIC} = -2L(\hat{\Theta}) + 2p$ \cite{sakamoto_akaike_1986}.

\subsection{Cluster adjustment in Huber sandwich estimator}

\citet{huber_behavior_1967} provided a consistent covariance matrix estimator for model parameters, known as `Huber sandwich estimator'. We list the essential derivation steps here; details can be found in \cite{freedman_so-called_2006}. 

Expanding the log-likelihood function $L(\Theta)$ in a Taylor series around true parameter $\Theta_0$, and given that $\hat{\Theta}$ is the solution of $L'(\Theta)=0$, we have:
\begin{equation*}
    \mathrm{cov} (\hat{\Theta}) = [-L''(\Theta_0)]^{-1}[\mathrm{cov} (L'(\Theta_0))][-L''(\Theta_0)]^{-1}.
\end{equation*}
By convention, the term $-L''(\Theta_0)$ is computed by summing up the Hessian matrices estimated from the sample and denoted by $\hat{J}$
\begin{equation}
    \hat{J} = -L''(\hat{\Theta}) = \sum_{i} \pdv{\log p(y_i|X_i,\hat{\Theta}) }{\theta}{\theta^{T}}
\end{equation}
Note that $\hat{J}/N$ converges to $J$ by the Law of Large Numbers.
The $\mathrm{cov} (L'(\hat{\Theta}))$ is estimated similarly as the sum of the Fisher information matrices and denoted by $\hat{K}$
\begin{equation}
     \hat{K} = \mathrm{cov} (L'(\hat{\Theta})) = \sum_{i} G_i^{T}G_i.
\end{equation}
where $G_i = \pdv{\log p(y_i|X_i,\hat{\Theta})}{\theta}$ denote the $1 \times p$ gradient vector on the $i^{\text{th}}$ observation. Using the sum of the covariance matrices of individual gradient vectors to compute the overall covariance matrix implies that the gradient vectors are independent across observations. The `Huber sandwich estimator' is given by:
\begin{equation*}
    \mathrm{cov}(\hat{\Theta}) = \hat{J}^{-1}\hat{K}\hat{J}^{-1}.
\end{equation*}    

When observations are correlated within clusters, it is invalid to assume independent gradient vectors across observations. Instead, the gradient vectors are correlated within each cluster, but the gradient vector sums for each cluster are assumed to be independent across clusters \cite{barlow_robust_1994, mccullagh_generalized_2019}. Thus, the following clustering adjustment was applied to $\hat{K}$ in equation (5) \cite{harrell_overview_2015, freedman_so-called_2006, barlow_robust_1994, mccullagh_generalized_2019}, to obtain a clustered Fisher information matrix estimator:
\begin{equation}
    \hat{K}_c = \sum_{j=1}^{M} [(\sum_{i=1}^{N_j}G_{ij})^{T}(\sum_{i=1}^{N_j}G_{ij})].
\end{equation} 
where $M$ is the number of clusters, $N_j$ is the number of observations in the $j^{\text{th}}$ cluster, $G_{ij}$ is the gradient vector for the $i^{\text{th}}$ observation in the $j^{\text{th}}$ cluster. Then, the variance-covariance matrix estimator for clustered data is given by 
\begin{equation*}
    \mathrm{cov}_c(\hat{\Theta}) = \hat{J}^{-1}\hat{K}_c\hat{J}^{-1}
\end{equation*}

\subsection{Mathematical formula of NICc}
Here, we extend the NIC to a clustered data setting. We substitute the Fisher information $K$ in the Network Information Criterion in equation (2) with the estimator $\hat{K}_c$ in equation (5), and defined the clustered estimator of the Network Information Criterion (NICc) as:
\begin{equation}
     \text{NICc} = -2L(\hat{\Theta}) + 2\text{trace}[\hat{J}^{-1}\hat{K}_c].
\end{equation}
Substituting $K$ with $\hat{K}$ in equation (4) leads to an observation-based standard estimator of the Network Information Criterion, which asymptotically approximates AIC when the model is true \cite{stone_asymptotic_1977, ripley_pattern_1996}. 

Hereafter, we will let \textbf{NIC} denote the standard estimator of the Network Information Criterion, and \textbf{NICc} denote the clustered estimator of the Network Information Criterion.

For clustered data with a large number of observations within each cluster, the NICc penalizes the log-likelihood of complex models more so than NIC. The $\hat{K}_c$ in NICc first sums up the gradient vectors of all observations within the same cluster and calculates a cluster-based Fisher information matrix for each cluster. Then, $\hat{K}_c$ is the summation of all cluster-based Fisher information matrices, rather than the summation of all Fisher information matrices of all observations.

\begin{figure}
    \centering
    \includegraphics[width=0.9\linewidth]{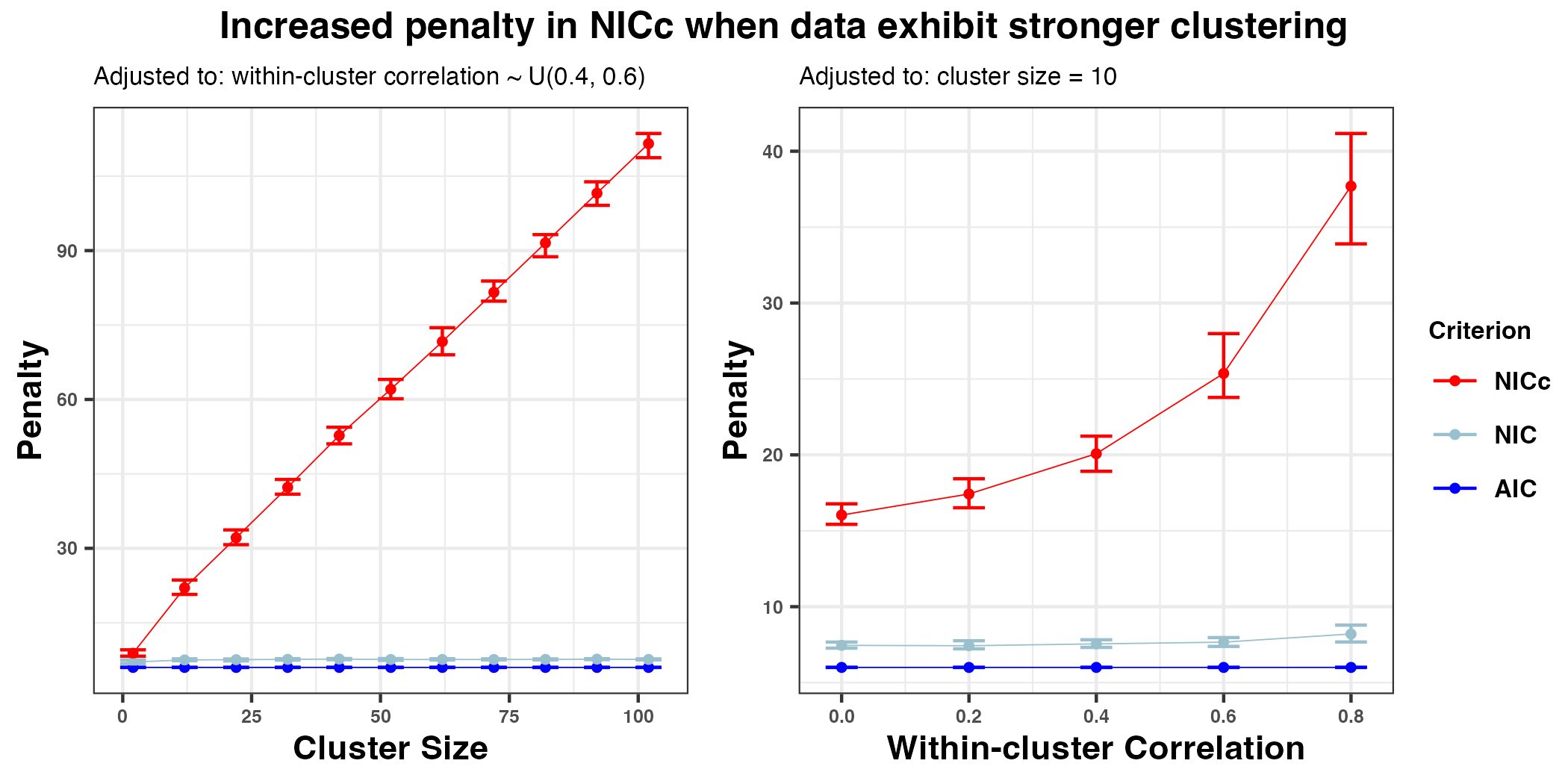}
    \caption{NICc more effectively prevents over-parameterized prediction models when data exhibit stronger clustering. Linear regression was used to develop prediction models on simulated clustered data with Gaussian responses. On the \textbf{left} panel, as the clusters contain a larger number of positively correlated observations, the penalty in NICc increases with cluster size, whereas NIC and AIC, defined for i.i.d. samples, remain unchanged. On the \textbf{right} panel, when clusters contain more positively correlated observations, NICc induces a larger penalty accordingly, while AIC and NIC remain almost the same. The within-cluster correlation is determined by the AR1 coefficient, which is sampled for each cluster from an Uniform distribution $\mathcal{U}(\textit{x}, \textit{x}+0.2)$ for $x = 0,0.2,0.4,0.6,0.8$.}
    \label{fig:penalty}
\end{figure}

Mathematically, for the $j^{\text{th}}$ cluster with $N_j$ observations, let $G_{ij}$ denote the $1 \times p$ gradient vector of the $i^{\text{th}}$ observation in the cluster. Let $g_{ij}(q)$ denote the $q^{\text{th}}$ element in $G_{ij}$, which is the gradient with respect to parameter $q$. The $q^{\text{th}}$ diagonal element in the observation-based Fisher information matrix $\hat{K}$ in NIC is:
\[
\hat{K}(q, q) = \sum_{i = 1}^{N_j} g_{ij}(q)^2,
\]
whereas in the cluster-based Fisher information matrix $\hat{K}_c$ in NICc, it is:
\[
\hat{K}_c(q, q) = [\sum_{i=1}^{N_j} g_{ij}(q)]^2.
\]
If observations within the same cluster are closer in value to each other (positively correlated \cite{mccullagh_generalized_2019}), the within-cluster gradient vectors are also positively correlated by having the same sign and similar values. Thus, a larger number of correlated observations within a cluster will result in a larger $\hat{K}_c(q, q)$ compared to $\hat{K}(q, q)$, leading to a higher penalty in NICc than in NIC. As the number of parameters $p$ increases in more complex models, the $\text{trace}[\hat{J}^{-1}\hat{K}_c]$ in NICc increases to a greater extent than the $\text{trace}[\hat{J}^{-1}\hat{K}]$ in NIC. As shown in Figure \ref{fig:penalty}, NICc imposes larger penalties compared to NIC and AIC when the number of positively correlated observations within clusters increases, and when the strength of correlation within clusters increases. Consequently, NICc can more effectively prevent over-parameterized prediction models developed on strongly clustered data compared to NIC and AIC.

\section{Simulation study} \label{sec_simu}

\subsection{Generating process for clustered data}
For clustered data, we used a generalized linear mixed effects model to generate Binomial or Gaussian response variable, as well as predictor variables that have fixed and random effects on the response. We sampled 50 clusters with $N_{j} \in \{2, 10, 50, 100, 150\}$ observations per cluster (denoted by $j$) and specified a total of $p \in \{5, 6, ..., 10\}$ predictor variables. All predictors had fixed effects and $80\%$ were set to have random effects. Each simulation condition was iterated 100 times.

For the Binomial distribution, the dichotomous response variable $y_{ij}$ for the $i^{\text{th}}$ observed vector of the $j^{\text{th}}$ cluster follows a Bernoulli distribution with probability $p_{ij}$:
\[f(y_{ij}; p_{ij}) =
  \begin{cases}
    p_{ij}     & \text{if $y_{ij} = 1$}, \\
    1 - p_{ij} & \text{if $y_{ij} = 0$};
  \end{cases} \;\;
\log( \frac{p_{ij}}{1-p_{ij}} ) = X_{ij}\beta + Z_{ij}b_j .
\]
For the Gaussian distribution, $y_{ij}$ follows a Gaussian distribution:
\[y_{ij} \sim \mathcal{N}(\mu_{ij}, 2), \;\;
\mu_{ij} = X_{ij}\beta + Z_{ij}b_j,\]
where $X_{ij}$ is the vector of predictor variables with fixed effects on the response variable, and $Z_{ij}$ contains predictor variables with within-cluster random effects on the response variable. $\beta$ is the coefficient vector for the fixed effects. $b_j$ is the coefficient vector for the random effects within the $j^{\text{th}}$ cluster. 
We generated $X_{ij}$, $\beta$ for the $k^{\text{th}}$ fixed-effect predictor, and $Z_{ij}$, $b_j$ for the $s^{\text{th}}$ random-effect predictor within the $j^{\text{th}}$ cluster, as follows: 
\[
X_{ij}(k) \sim \mathcal{N}(0, 1),
\]
\[
\beta(k) \sim \mathcal{N}(0, \sigma_{\beta} = 5).
\]


\[
Z_{ij}(s) \sim AR1(\phi_i (s), \omega=1), \;\;
\phi_i(s) \sim \mathcal{U}(\phi, \phi+0.2), \;\; \phi \in \{0, 0.4, 0.8\}.
\]


\[
b_j(s) \sim \mathcal{N}(0, r_{b}\sigma_{\beta} ), \;\;  r_{b} \in \{0.5, 1, 10\}.
\]


In our design, three factors controlled the strength of clustering in the generated data: the number of observations per cluster ($N_{j}$), the ratio of standard deviation of random-effect coefficients to fixed-effect coefficients ($r_{b}$), and the AR1 autoregressive coefficient level ($\phi$) per cluster per predictor. 

\begin{enumerate}
    \item Given \( r_b > 0 \) and \( \phi > 0 \), a larger volume of correlated observations per cluster (\( N_j \)) will result in stronger clustering of data. 
    \item Given \( N_j > 1 \) and \( \phi > 0 \), relatively larger random-effect coefficients (larger \( r_b \)) will generate more correlated observations within each cluster in the response variable, leading to stronger clustering in the data due to the response.
    \item Given \( N_j > 1 \) and \( r_b > 0 \), a higher level of the autoregressive coefficient \( \phi \) in the AR1 process will create more correlated observations within each cluster in the predictor variables, resulting in stronger clustering in the data due to the predictors.
\end{enumerate}

For prediction model selection, both 
under-parameterized and over-parameterized prediction models should exhibit worse out-of-cluster performance, as determined by leave-one-cluster-out deviance. For data simulated under a given clustering condition defined above, we started with a total of 5 predictors $(p=5)$, and created overfitting scenarios by adding fifth-order polynomial-transformed terms of random-effect predictors. This transformation increases the complexity of the random-effect variables, making the model more likely to overfit individual clusters.

\subsection{Leave-one-cluster-out cross-validation by NICc}

\subsubsection{Approximating looDeviance}

We used linear and logistic regression to model synthetic clustered data with Gaussian and binomial responses, respectively. We controlled simulation conditions by the number of observations per cluster $N_{j}$, the total number of predictor parameters $p$, the strength of the random effects relative to the fixed effects $r_{b}$, and the autoregressive correlation in predictor variables $\phi$. Under varying conditions, we compared the accuracy of NICc, NIC, AIC and BIC in approximating out-of-cluster performance determined by looDeviance. Specifically, for 100 iterations under each simulation condition, we measured the approximation error of each criterion by the deviance between the criterion and looDeviance, normalized by total number of observations:

\[\text{Error} = \frac{\text{IC - looDeviance}}{\text{Total number of observations} }\]

We assessed how each criterion's approximation to looDeviance was influenced by the three factors that controlled the clustering strength in the data: number of observations per cluster ($N_{j}$), the relative strength of random effects ($r_{b}$), and the autoregressive coefficient in predictors ($\phi$). Specifically, we evaluated each factor's marginal impact on a criterion's approximation to looDeviance by varying that factor while keeping the remaining two factors fixed at their median levels ($N_j=50, r_b=1, \phi = 0.4$), which preserved a moderate clustering structure in the data.

We extended the simulation to examine the accuracy of NICc in approximating looDeviance under various special scenarios, which included K-fold cross-validation, unbalanced cluster sizes, small samples, and rare events. Each scenario was tested under weak, moderate, and strong clustering conditions. If not otherwise specified: weak clustering is determined by \(N_{j} = 10\), \(r_{b} = 0.5\), and \(\phi = 0\); moderate clustering is determined by \(N_{j} = 50\), \(r_{b} = 1\), and \(\phi = 0.4\); and strong clustering is determined by \(N_{j} = 100\), \(r_{b} = 10\), and \(\phi = 0.8\). 

The special scenarios are generated as follows. First, K-fold cross-validation is commonly used in practice, when the number of observed clusters are extensive. We extended the number of clusters to 100 and examined the accuracy of NICc in approximating the K-fold deviance with $K \in \{10, 50, 80\}$. Second, for unbalanced cluster sizes, we compared each criterion's approximation accuracy for data with balanced versus unbalanced cluster sizes. The unbalanced cluster sizes were created by randomly keeping 20\% to 80\% of observations per cluster. For balanced cluster sizes, \(N_j\) was set to be half that of corresponding unbalanced conditions to maintain the same average number of observations per cluster.
Third, for small samples, the sandwich variance estimator in generalized estimating equation methods can be biased towards the direction of zero, underestimating the true variance \cite{pan_small-sample_2002, li_small_2015}. We extended the total number of clusters to 10 and examined the impact of this small number of clusters on the approximation accuracy of NICc, which is based on the concept of the clustered robust sandwich estimator. For weak clustering, the lowest number of observations per cluster is set to 25 to ensure a sufficient total number of observations for the prediction models. 
Lastly, for rare events of binomial responses, the prevalence (ratio of clusters containing 1 against the total number of clusters) in the simulated data is controlled to be around 0.15 for each iteration.

\subsubsection{Prediction model selection}


For each criterion (looDeviance, NICc, NIC, AIC or BIC), we created a set of candidate models by incrementally adding a predictor variable to the existing model (start with a model containing one predictor) that resulted in the greatest reduction in the criterion, until all predictors were included. We then identified two types of optimal models selected by each criterion: the model size that minimized the criterion or the smaller model size that was one standard error (1SE) away from the minimum, should simpler models be preferred according to Occam's razor \cite{young_simplicity_1996}. 

We compared the performance of NICc, NIC, AIC and BIC in selecting the optimal model, as determined by looDeviance, with respect to model size and model misspecification. First, we measured each criterion's error in estimating the optimal model size determined by ground-truth looDeviance, by comparing the difference in model sizes selected by each criterion to the model size selected by looDeviance. Second, we assessed each criterion's model misspecification by calculating the Jaccard index \cite{real_probabilistic_1996} between the set of predictors in the optimal model selected by looDeviance, and the set of predictors in the optimal model selected by each criterion. A higher Jaccard index indicates less model misspecification of a criterion compared to looDeviance.

\subsection{Out-of-cluster performances of GLM, GEE and RE}

Although not the main focus of this article, we used a simulation study to replicate prior results, as well as extend it to more generalized conditions, that standard GLMs exhibit non-inferior out-of-cluster predictive performance and high computational efficiency compared to GEE and RE methods, when developing prediction models for clustered data with Gaussian or binomial responses. Note that when using GLM methods, such as linear and logistic regression, the clustered Huber sandwich adjustment to the standard errors constitutes a special case of specifying an independent working correlation structure in corresponding GEE methods \cite{liang_longitudinal_1986}. 


We compared three methods for developing prediction models using the simulated clustered data with Gaussian and binomial responses: 1) regular linear and logistic regression; 2) GEE methods with an AR1 correlation structure, and 3) RE methods that used the same fixed and random effects specifications as the data generating process. We calculated the looDeviance of each method under each simulation condition for 100 iterations. For the RE method, out-of-cluster prediction was achieved by using the fixed effects estimated from all but one cluster, to make predictions on the left-out cluster. We also calculated leave-one-cluster-out mean squared error (looMSE) for Gaussian responses and leave-one-cluster-out area under the receiver operating characteristic curve (looAUC) for binomial responses to measure out-of-cluster performance. We evaluated the time (seconds) consumed by each method for conducting leave-one-cluster-out cross-validation, the non-converging rate of each method over 100 iterations, and the out-of-cluster performances of each method under varying strengths of clustering as well as special scenarios, including small samples, unbalanced sample sizes, and rare events.

\section{Simulation results} \label{sec_simu_res}

\subsection{Leave-one-cluster-out cross-validation by NICc}
AIC and NIC had nearly identical values across all simulation conditions, supporting the finding that NIC is the asymptotic form of the leave-one-observation-out deviance of a model, and AIC is asymptotically equivalent to the NIC when the model is true. Therefore, the following simulation results for the AIC can also apply to the NIC. 

\subsubsection{Accuracy in approximating leave-one-cluster-out deviance}

\begin{figure}[htp]
    \centering
    \makebox[\textwidth]{\includegraphics[width=\textwidth]{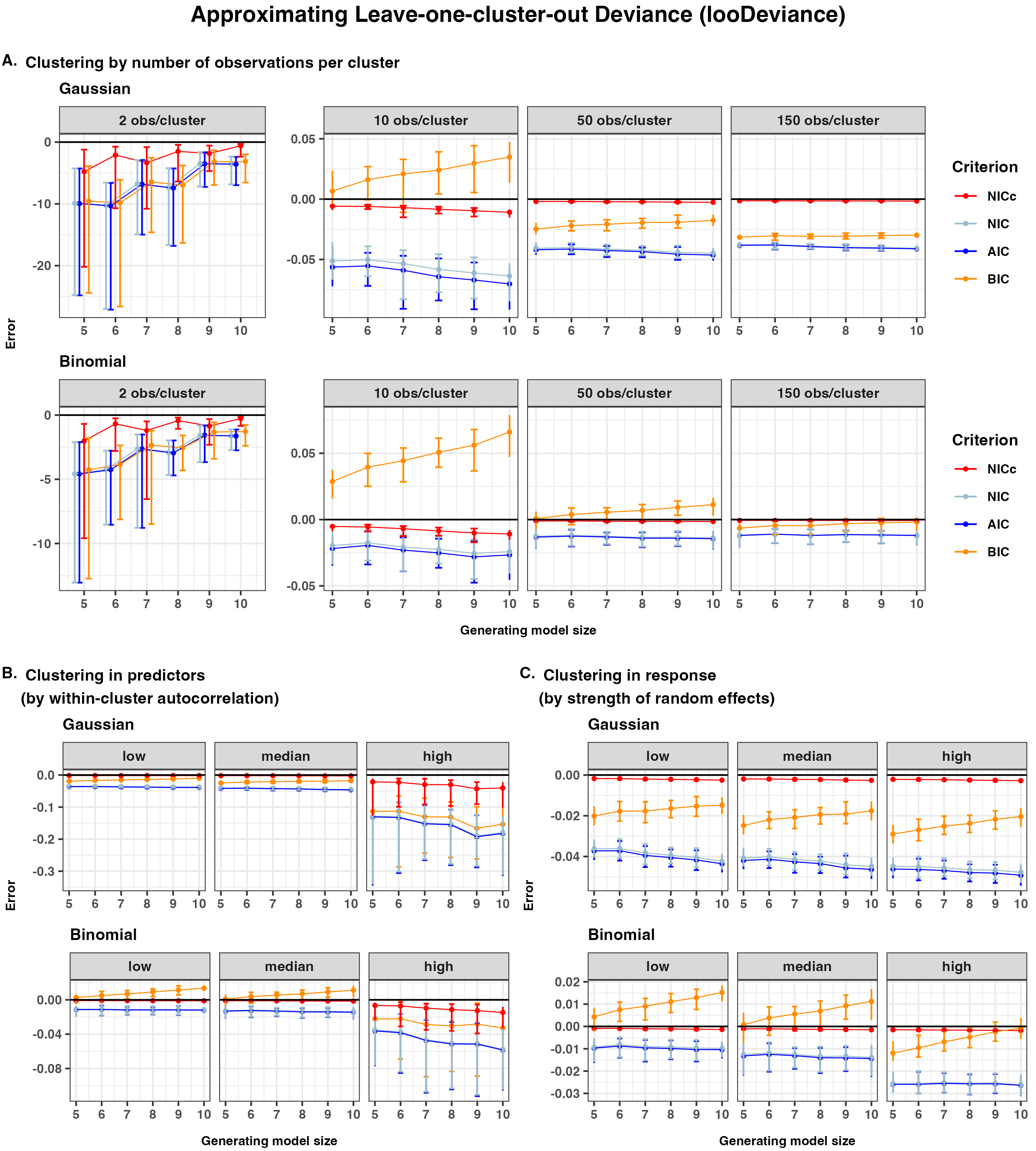}}
\caption{Accuracy of approximating leave-one-cluster-out deviance (looDeviance) by NICc, NIC, AIC, and BIC for clustered data with Gaussian or binomial responses. Colors indicate different information criteria, with the black horizontal line representing looDeviance. For each tile, the x-axis represents the total number of predictors (generating model size), and the y-axis represents the difference between each criterion and looDeviance, normalized by the total number of observations. The points correspond to the median value over 100 iterations, and error bars represent the 2.5th and 97.5th percentiles. \textbf{Panel A} illustrates the impact of the number of observations per cluster on each criterion's approximation to looDeviance, with a median level of clustering in predictors ($\phi=0.4$) and response ($r_b=1$). \textbf{Panel B} shows the impact of clustering in predictors on each criterion's approximation to looDeviance, given 50 observations per cluster and a median level of clustering in response ($r_b=1$). Low, median, or high clustering in predictors is controlled by AR1 autoregressive coefficients ($\phi \in {0, 0.4, 0.8}$). \textbf{Panel C} demonstrates the impact of clustering in the response variable on each criterion's approximation to looDeviance, given 50 observations per cluster and a median level of clustering in predictors ($\phi=0.4$). Low, median, or high clustering in response is controlled by the ratio of random effects to fixed effects ($r_b \in {0.5, 1, 10}$).}\label{fig_nic_vs_aic}
\end{figure}

As shown in Figure \ref{fig_nic_vs_aic}, when approximating out-of-cluster predictive performance for clustered data, NICc exhibited good accuracy and lower approximation error than AIC and BIC under all simulation conditions. Clusters with two observations commonly arise in biologically linked experimental units, such as eyes from the same patient or sets of twins \cite{durkalski_analysis_2003}. Though all criteria exhibited higher error with only two observations per cluster, the NICc maintained notably higher accuracy than AIC and BIC. As shown in panels B and C, stronger clustering led to a greater approximation error in AIC and BIC, but not NICc, wherein the NICc remained accurate in approximating looDeviance. The BIC occasionally exhibited similar approximation accuracy to the NICc, which is due to the occasional equality between the log of the total number of observations times the total number of parameters and the penalty in NICc under certain simulation conditions. Under general conditions, BIC showed a greater approximation error than NICc.

\begin{figure}[htp]
    \centering
    \makebox[\textwidth]{\includegraphics[width=\textwidth]{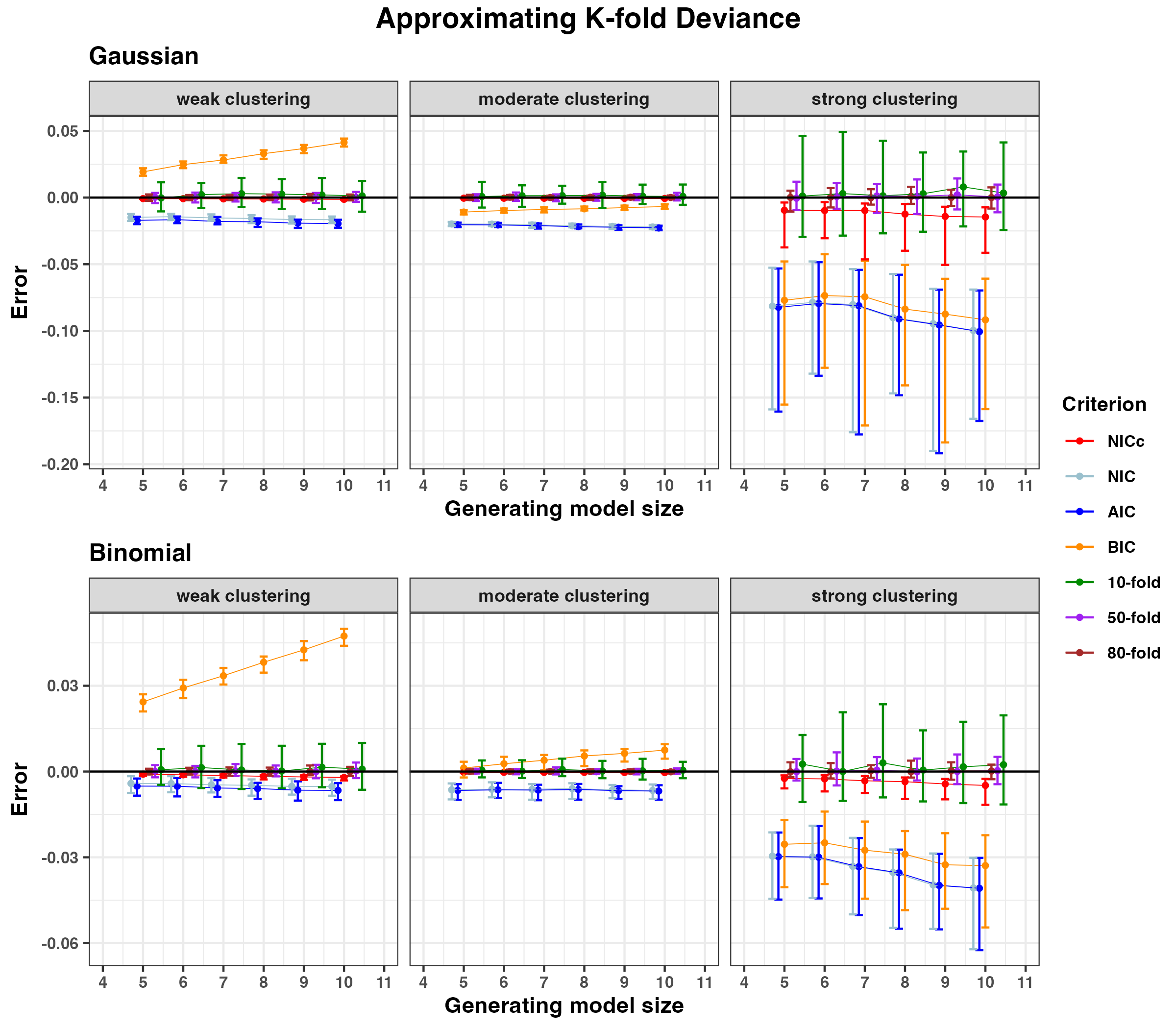}}
\caption{Accuracy of approximating K-fold deviance (splitting clusters) by NICc, NIC, AIC, and BIC for data containing 100 clusters. Linear and logistic regression were used for Gaussian and binomial responses respectively. Colors indicate different information criteria and K-fold deviance with K equal to 10, 50, and 80. The black horizontal line refers to looDeviance. For each tile, the x-axis represents the total number of predictors corresponding to the generating model size, and the y-axis represents difference between the looDeviance and each criterion or K-fold Deviance, normalized by the total number of observations. The error bar around each point represents the 2.5th and 97.5th percentiles. }\label{fig_k_fold}
\end{figure}

Additionally, as shown in Figure \ref{fig_k_fold}, the NICc can successfully approximate K-fold cross-validated deviance with K ranging from a relatively small number (e.g., 10) to a larger number (e.g., 80). In contrast, the AIC and BIC consistently performed worse than the NICc. Simulation also showed that the NICc exhibited higher approximation accuracy than the AIC and BIC, when strongly clustered data had unbalanced cluster sizes or a small number of clusters (Figure \ref{fig_specfial_scenario}). 
When the binomial response contains rare events, both NICc and AIC showed higher approximation accuracy than BIC for varying clustering strengths.

\subsubsection{Accuracy in prediction model selection}

\begin{figure}[htp]
    \centering
    \makebox[0.95\textwidth]{\includegraphics[width=0.95\textwidth]{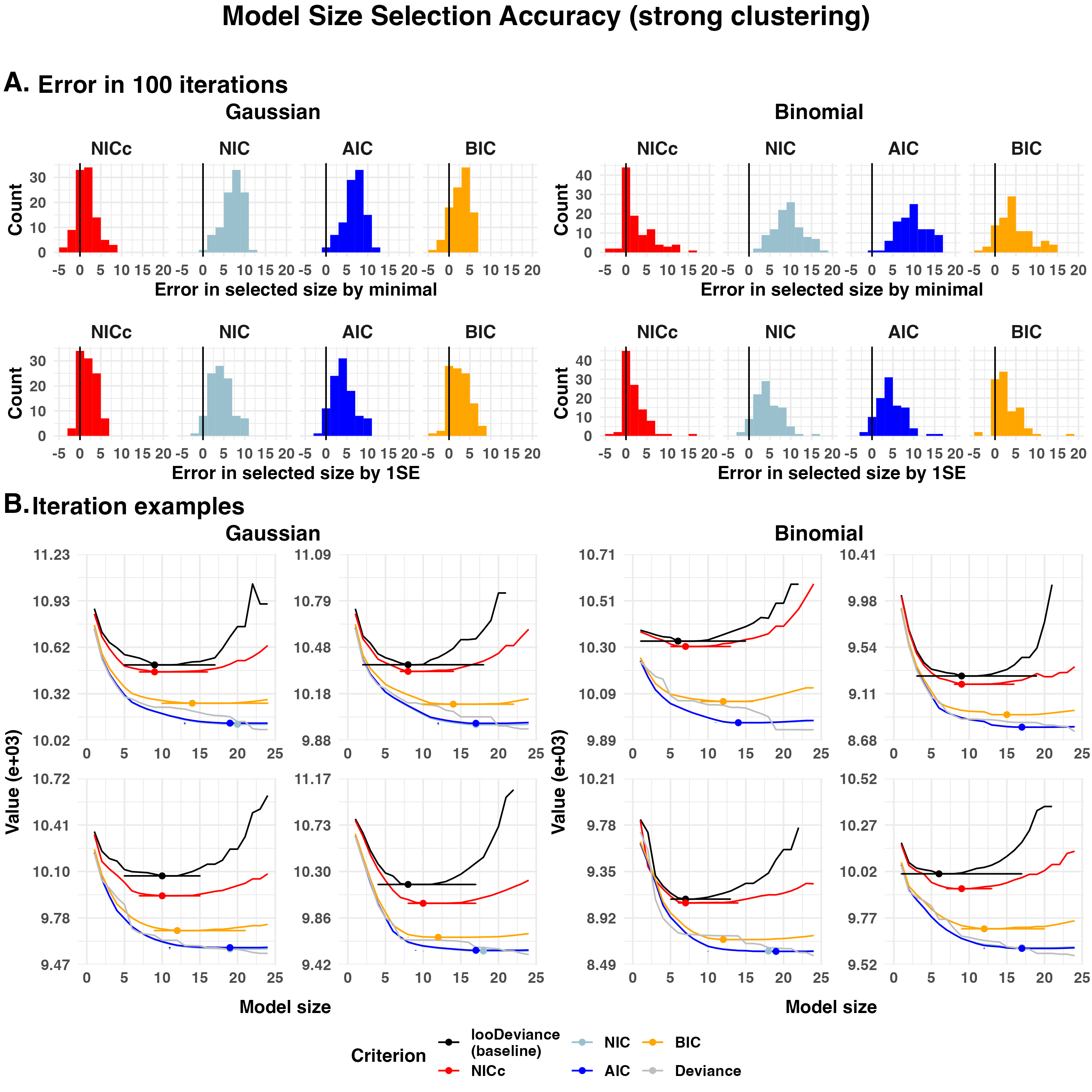}}
\caption{Accuracy of model size selected by NICc, NIC, AIC and BIC when data exhibit strong clustering. Linear regression and logistic regression are used for Gaussian and binomial responses, respectively.  \textbf{Panel A} shows the accuracy of NICc, NIC, AIC or BIC in selecting the optimal model size at its minimal value or at the smaller model at one standard error (1SE) from the minimum, across 100 iterations. The error in model size is measured by the difference in the model size selected by each criteria compared to the model size selected by looDeviance (baseline). \textbf{Panel B} depicts four examples from the 100 iterations, where the curved line represents the criterion value at each model size. The point marks its minimum value with the corresponding model size, and the horizontal line (error bar) indicates the set of model sizes within 1SE of the minimum.}\label{fig_model_select_strong}
\end{figure}

\begin{figure}[htp]
    \centering
    \makebox[0.9\textwidth]{\includegraphics[width=0.9\textwidth]{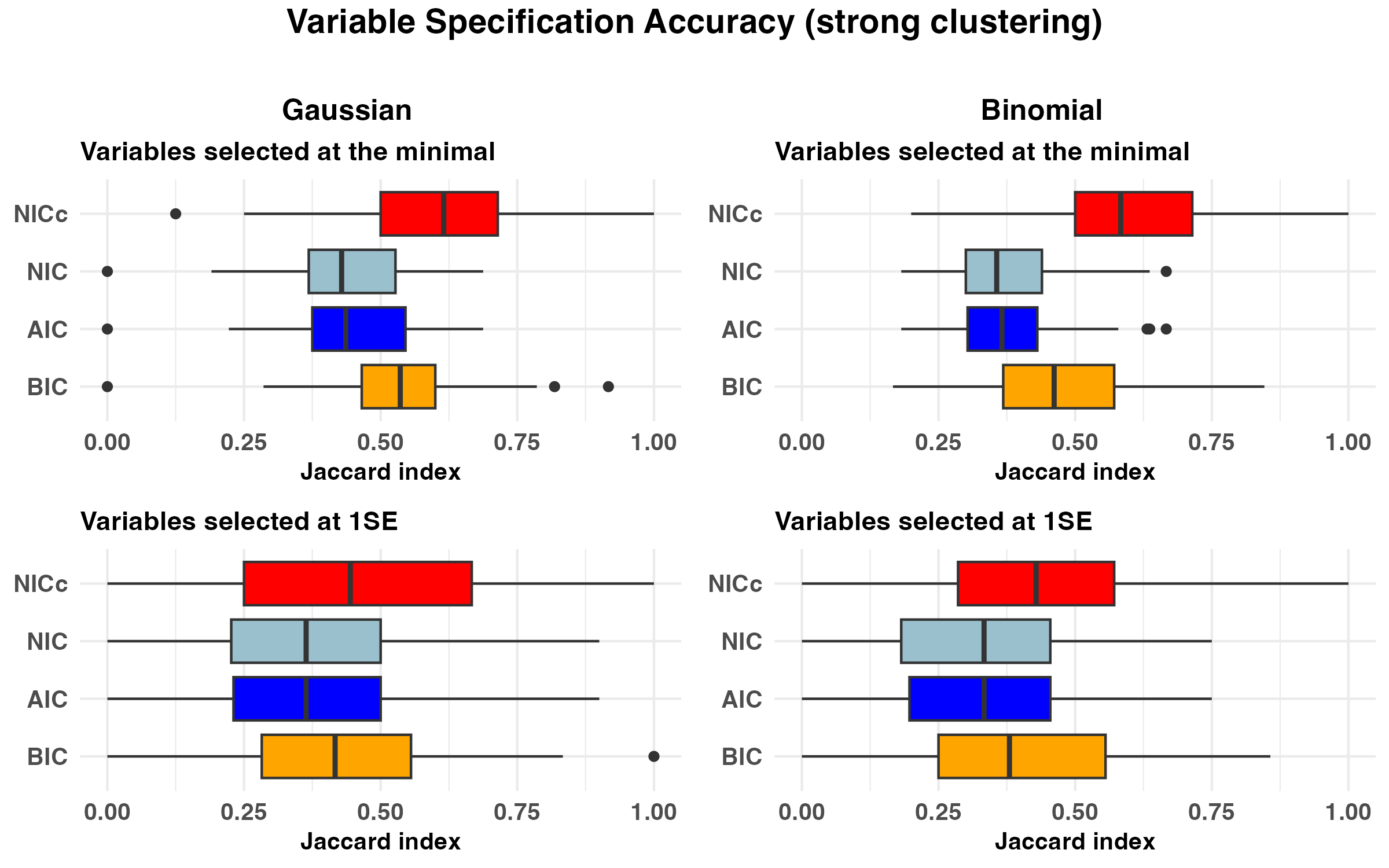}}
\caption{Accuracy of variable specification in optimal models selected by NICc, NIC, AIC and BIC for data under strong clustering condition. Linear regression and logistic regression are used for Gaussian and binomial responses respectively. The accuracy of variable specification is measured using the Jaccard index between the set of variables selected by each criterion and the set of variables selected by leave-one-cluster-out deviance (looDeviance). The Jaccard index ranges from 0 to 1, with a higher index indicating greater accuracy in the set of variables specified by a criterion.}\label{fig_model_misspecification}
\end{figure}

When data exhibited relatively strong clustering, NICc had higher accuracy in selecting the optimal predictive model size, compared to AIC and BIC. As shown in Figure \ref{fig_model_select_strong} Panel A, for both the optimal model and at a 1SE lower than a criterion's minimum, the NICc had near 0 error, also lower compared with the other criteria. 
As illustrated in the iteration examples in Panel B, NICc imposed stronger penalties on over-parameterized models compared to the AIC and BIC. Therefore, the NICc aligned with the increase in looDeviance at larger model sizes, whereas AIC and BIC tended to remain flat or continue decreasing, indicating that the NICc prevented overfitting more effectively than the AIC and BIC. Additionally, the NICc demonstrated better variable selection accuracy compared to AIC and BIC. As shown in Figure \ref{fig_model_misspecification}: the Jaccard index between the optimal model selected by NICc and the optimal model selected by looDeviance was closer to 1 than the other criteria. 
When the clustering was weak or moderate, as shown in Figure \ref{fig_model_select_not_strong}, the model size selection accuracy and variable specification accuracy of NICc remained the highest amongst all criteria. The AIC outperformed the BIC under weaker clustering conditions and exhibited similar accuracy as NICc. 

\subsection{Out-of-cluster performances of GLM, GEE and RE}

As shown in Figures \ref{ooc_performance} and \ref{ooc_performance_special}, across all simulation conditions, the linear and logistic regression (GLM) exhibited almost identical out-of-cluster predictive performance as the GEE method and the RE method. Under special scenarios of unbalanced cluster size, small number of clusters, and rare events of the binomial response, the GLM also demonstrated non-inferior population-level prediction compared to RE and GEE methods. This result is not unexpected because only the population-averaged effects from RE or GEE were used in prediction on unseen clusters. 

Nevertheless, GLM and GEE methods were much more computationally efficient than the RE method under almost all conditions. The RE methods generally took over 50 times longer than other methods for Gaussian responses and over 1000 times longer for binomial responses. GEE occasionally required more time than GLM to estimate the autoregressive coefficient. Furthermore, the non-convergence rate of the RE method was notably higher than that of other methods, especially when data exhibited unbalanced cluster sizes, small number of clusters, and/or lower event prevalence.

Take together, these results suggest that regular linear or logistic regression has non-inferior performance along with higher convergence rates and computational efficiency compared to RE and GEE methods when developing population-level prediction models on clustered data with Gaussian or binomial responses. In practice, when applying the RE method to massive datasets with high-dimensional feature spaces becomes computationally infeasible, it is advisable to use appropriate imputation techniques as well as techniques to address extreme values before implementing GLM methods.

\section{Empirical Example} \label{sec_empirical}

To illustrate our proposed criteria, we used a published dataset of $N=65353$ daily observations from $M=2964$ infants from the University of Virginia Neonatal Intensive Care Unit (NICU) between 2012 and 2016 \cite{emp_data}. We developed a logistic regression model to predict the binary event of death within the next 7 days. The event is rare with only $385$ events (rate=$385/63353=0.006$). The features included both the patient at-birth demographics and summary statistics from one daily 10-minute record of heart rate (HR) and oxygen saturation (SPO2) vital signs. Demographic variables are unique per infant, but vital signs are available each day of the stay, which average about $22$ repeated measures per infant.
Patient demographics and variable descriptions are provided in Table \ref{table_demo} and Table \ref{table_vars}.

\begin{table}
\centering
\caption{Empirical Example: Patient Demographics}
\centering
\begin{tabular}[t]{llll}
\toprule
  & Overall & Survived & Died\\
N & 2964 & 2846 & 118\\
\midrule
Race (\%) &  &  & \\
Caucasian & 2135 (72.2) & 2046 (72.1) & 89 (75.4)\\
Black & 552 (18.7) & 534 (18.8) & 18 (15.3)\\
Other & 270 ( 9.1) & 259 ( 9.1) & 11 ( 9.3)\\
\addlinespace
Gender = Male (\%) & 1696 (57.2) & 1623 (57.0) & 73 (61.9)\\
Ethnicity = Hispanic (\%) & 113 ( 3.8) & 109 ( 3.8) & 4 ( 3.4)\\
Multiple = Yes (\%) & 429 (14.5) & 420 (14.8) & 9 ( 7.6)\\
Gestational age in weeks (mean (SD)) & 35.18 (4.45) & 35.26 (4.36) & 33.26 (5.92)\\
Birth weight in grams (mean (SD)) & 2495.97 (995.10) & 2511.59 (984.27) & 2116.09 (1171.65)\\
\bottomrule
\end{tabular}\label{table_demo}
\end{table}

\begin{table}

\caption{Empirical Example: Variable Description}
\centering
\begin{tabular}[t]{ll}
\toprule
Name & Description\\
\midrule
hr\_mean & Heart rate mean\\
hr\_std & Heart rate standard deviation\\
hr\_max & Heart rate maximum\\
hr\_min & Heart rate minimum\\
sp\_mean & SPO2 mean\\
\addlinespace
sp\_std & SPO2 standard deviation\\
sp\_max & SPO2 maximum\\
sp\_min & SPO2 minimum\\
EGA & Estimated gestational age in weeks\\
BWT & Birth weight in grams\\
\addlinespace
Apgar1 & Apgar score at 1-minute\\
Apgar5 & Apgar score at 5-minute\\
Vaginal & Vaginal delivery\\
C-section & C-section delivery\\
Steroids & Antenatal steroids\\
\addlinespace
InBorn & Born in hospital\\
BirthHC & Head circumference\\
Multiple & Multiple births\\
MaternalAge & Maternal age in years\\
\bottomrule
\end{tabular}\label{table_vars}
\end{table}

\begin{figure}[htp]
    \centering
    \makebox[\textwidth]{\includegraphics[width=\textwidth]{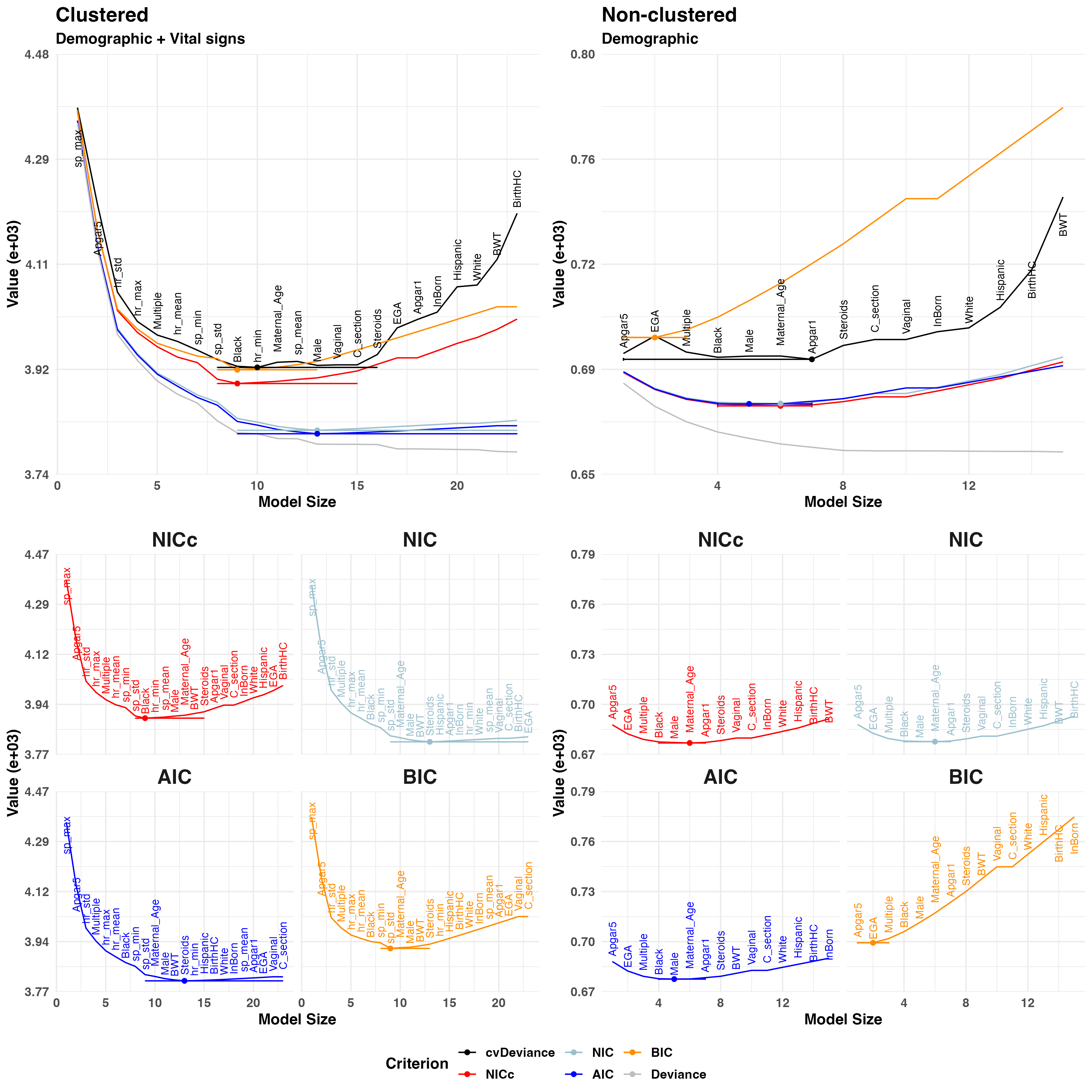}}
\caption{Empirical example of using NICc, AIC and BIC to select logistic regression models for predicting mortality in neonatal intensive care unit (NICU), using patient demographics (unique observation per patient) and vital signs (multiple observations per patient). 
}\label{fig_case_mortality_nicu}
\end{figure}

We developed two types of predictive models to predict NICU mortality: one used demographic variables as predictors, and the other used both demographics and vital sign statistics. Since the empirical dataset contained 2,964 patient clusters, which was too many to conduct leave-one-cluster-out cross-validation, we used cluster-based 100-fold cross-validated deviance (cvDeviance) to approximate out-of-cluster performance instead. For each criterion (cvDeviance, NICc, NIC, AIC, or BIC), we created a set of candidate models by incrementally adding a predictor variable to the existing model (starting with a model of one predictor) that resulted in the greatest reduction in the criterion until all predictors were included. In Figure \ref{fig_case_mortality_nicu}, we plotted each criterion's value and its selected variables at a given model size.

In the model that combined demographics and vital signs for clustered data, both NICc and BIC approximated cvDeviance well at each model size and indicated that a locally optimal model may include 8 to 15 variables, which was close to the model selection results by cvDeviance. AIC and NIC overestimated out-of-cluster performance for larger models, suggesting larger model sizes as candidates for the final model. This is consistent with our findings from the simulation study that demonstrated that the AIC fails to penalize models under strong clustering, resulting in over-parameterized models. Furthermore, the top 9 variables selected by minimizing NICc were identical and in the same order as those chosen by cvDeviance. The top 9 variables minimizing BIC were also identical to those selected by cvDeviance, though the variables `sp\_std' and `Black' could be misspecified if simpler models are preferred. AIC and NIC selected more misspecified variables than NICc and BIC. In the demographic-only model for i.i.d data, NICc was identical to AIC and NIC, while BIC penalized larger model sizes much more heavily than cvDeviance and other criteria.

\section{Discussion} \label{sec_discuss}

In biomedical settings, clustered data are commonplace, usually massive, and contain high-dimensional feature space. In this study, we aimed to improve the computational efficiency in developing prediction models for clustered data, with a focus on achieving high predictive performance on unseen clusters (out-of-cluster performance). These prediction models typically do not account for between-cluster heterogeneity during estimation via models like the GEE or GLMM due to the high computational burdens. Here, we introduced a clustered estimator of the Network Information Criterion (NICc) that can approximate the leave-one-cluster-out deviance (looDeviance) of such prediction models with twice-differentiable pseudo log-likelihood functions when the i.i.d assumption was violated for clustered data. 
We derived NICc by substituting the Fisher information matrix in the Network Information Criterion (NIC) with a clustering-adjusted estimator. 
The NICc imposes a greater penalty when the data exhibits stronger clustering, thereby allowing the NICc to better prevent over-parameterization. 

Both NICc and NIC are generalizable to parametric models with twice-differentiable log-likelihood functions. The choice of parametric model is not limited to a logistic regression or linear regression. An obvious strength of using NICc to develop predictive models on massive clustered data with high-dimensional features is the computational efficiency. When developing predictive models for clustered data with $N$ clusters, NICc can identify locally optimal models without the need for cross-validation. This makes it $N$ times faster than leave-one-cluster-out cross-validation and $K$ times faster than cluster-based K-fold cross-validation. Additionally, when the candidate models contain complex transformations of the observed predictors, the NICc can quickly determine whether a variable transformation improves out-of-cluster performance. However, we observe that for small sample sizes, the accuracy of NICc in approximating looDeviance deteriorates, although it remains higher than that of AIC and BIC. This is because the NICc is expected to perform better for larger clusters with larger sample sizes, similar to the asymptotic nature of NIC derived by \citet{stone_asymptotic_1977}. Therefore, correcting NICc for small sample sizes, similar to the corrected AIC (AICc) \cite{sugiura_further_1978}, warrants further investigation.

In a simulation study, we found that:

\begin{enumerate}
    \item NICc well approximates the out-of-cluster performance, as determined by looDeviance or K-fold clustered-based cross-validated deviance (cvDeviance), with a higher accuracy compared to AIC and BIC for varying strengths of clustering. 
    
    \item NICc leads to more accurate model selection than AIC and BIC and more accurate variable specification determined by looDeviance. When selecting prediction models for data with strong clustering, the NICc prevented over-parameterization more effectively than the AIC and BIC. 
    
\end{enumerate}
In general, NICc performed well in approximating out-of-cluster performance and selecting local optimal prediction models for data under varying simulation conditions. Considering unbalanced cluster sizes, the NICc demonstrates the same accuracy in approximating looDeviance as it does for balanced cluster sizes. For small samples (either cluster size or number of clusters), NICc had similar performance as the AIC and BIC for small samples with weak clustering, but it is better-suited for small samples with strong clustering. When modeling rare events, the NICc had similar accuracy as the AIC and the i.i.d estimator of NIC in approximating looDeviance. 

Other types of extensions based on NIC should be considered for further analysis. Nested data can exhibit hierarchical clustering structures, such as repeated measures of patients from multiple medical centers or longitudinal data where only measurements within a certain time window are considered correlated for each patient. For example, in NICU data, it might be reasonable to assume that measurements from the same infant taken more than a week apart are no longer correlated. Extending NICc to accommodate nested clustering structures, such as using a nested clustering adjustment on the Fisher information matrix in NIC, may offer potential benefits for cross-validation procedures designed for nested data. Another application would be to apply approach to estimating cross-validate performance using other common objective functions used in machine learning like sum of squared errors. The only requirement would be that the objective function is twice differentiable and can be represented well locally using quadratic Taylor series expansion.

Another contribution of this study involves comparing the out-of-cluster predictive performance of standard regression models with more complex methods that address cluster heterogeneity in model parameterization, namely the random effects (RE) method and the generalized estimating equations (GEE) method. We validated previous findings of the non-inferior out-of-cluster prediction performance of standard logistic regression compared to the random intercept model \cite{bouwmeester_prediction_2013} and extended this comparison to a broader range of clustering conditions, including varying clustering strengths, unbalanced cluster sizes, small numbers of clusters, and rare events. We found that standard linear and logistic regression demonstrated non-inferior out-of-cluster predictive performance, but much higher computational efficiency and convergence rates, than RE and GEE (with AR1 correlation structure) methods.
Other types of within-cluster correlation structures and other response variable distributions should be explored in further studies for a more comprehensive comparison between these methods.

\begin{appendices}
\section{Supplementary Figures}\label{secA1}
\begin{figure}[htp]
    \centering
    \makebox[\textwidth]{\includegraphics[width=\textwidth]{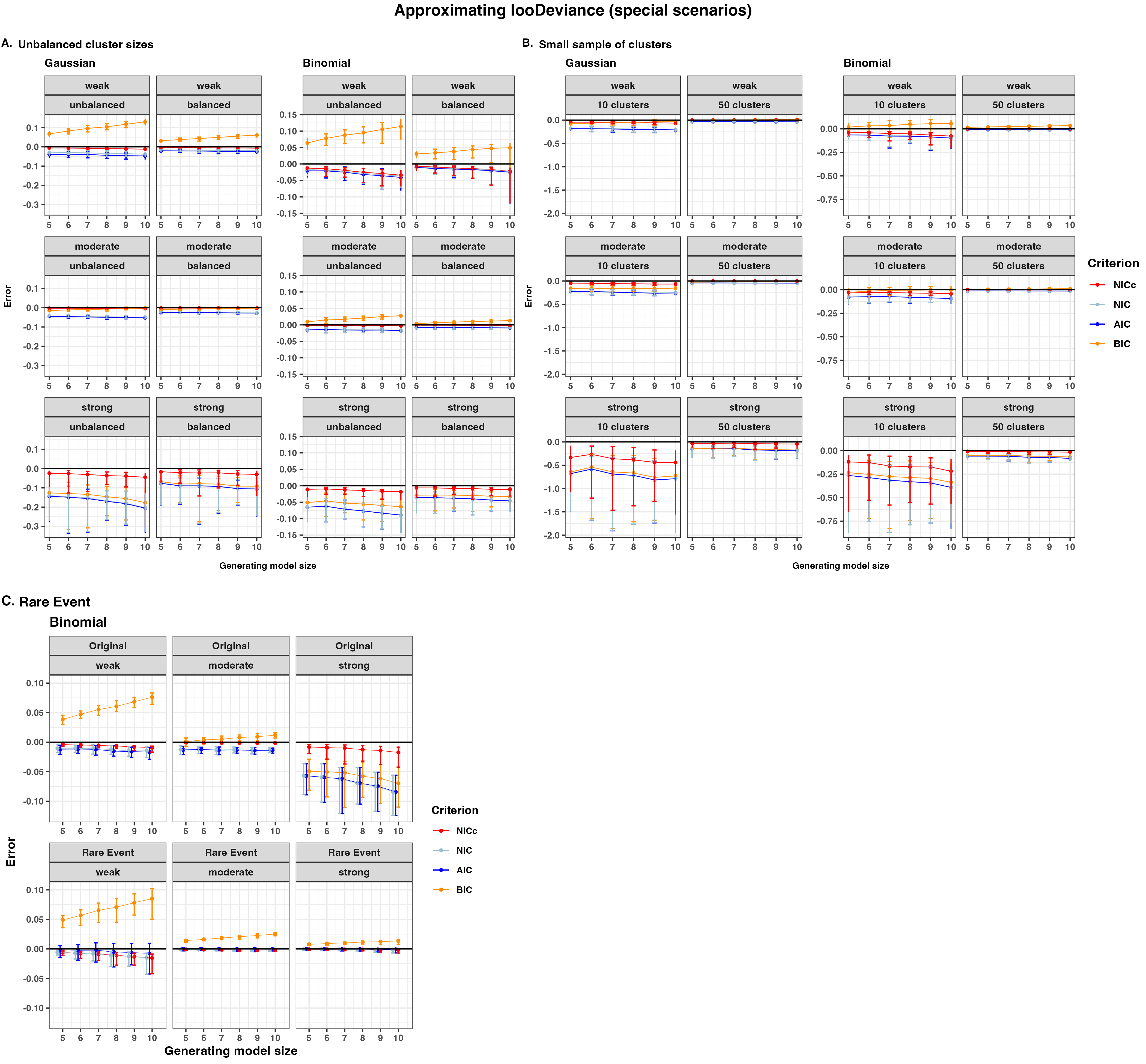}}
\caption{Accuracy of approximating leave-one-cluster-out deviance (looDeviance) by NICc, NIC, AIC, and BIC under weak, moderate and strong clustering conditions, in three special scenarios: balanced versus unbalanced cluster sizes (Panel A), small sample sizes (Panel B), and rare events (Panel C). Linear and logistic regression were used for Gaussian and binomial responses, respectively. In each tile, the x-axis represents the total number of predictors (generating model size), and the y-axis represents the difference between each criterion and looDeviance, normalized by the total number of observations.}\label{fig_specfial_scenario}
\end{figure}

\begin{figure}[htp]
    \centering
    \makebox[\textwidth]{\includegraphics[width=\textwidth]{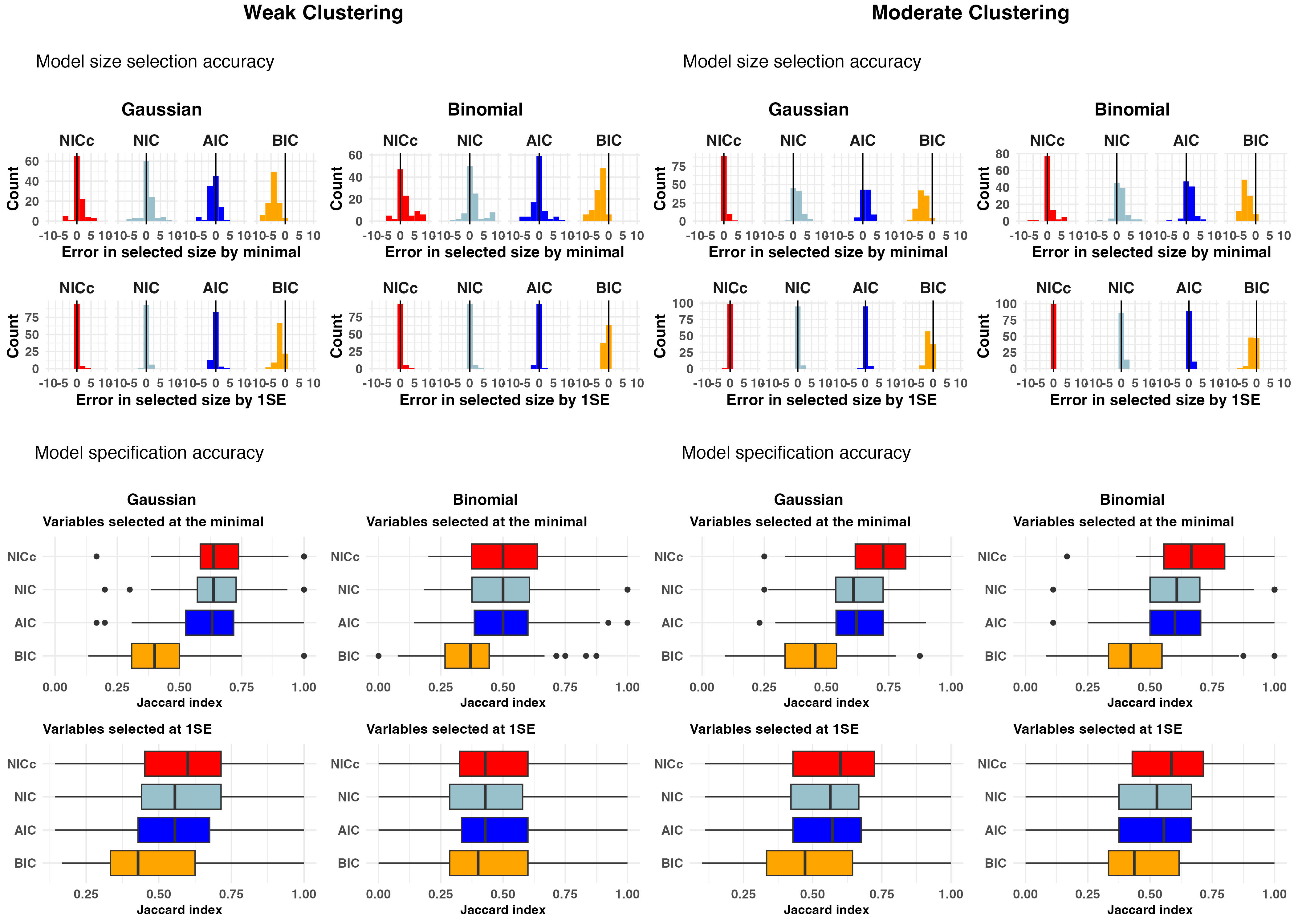}}
\caption{Model selection accuracy under weak and moderate clustering conditions. Linear regression and logistic regression are used for Gaussian and binomial responses respectively. Model size selection error is measured by the difference in the model size selected by each criteria compared to that by looDeviance (baseline). Variable specification accuracy is measured using the Jaccard index between the set of variables selected by each criterion and the set of variables selected by leave-one-cluster-out deviance (looDeviance). }\label{fig_model_select_not_strong}
\end{figure}
\begin{figure}[htp]
    \centering
     \makebox[\textwidth]{\includegraphics[width=\textwidth]{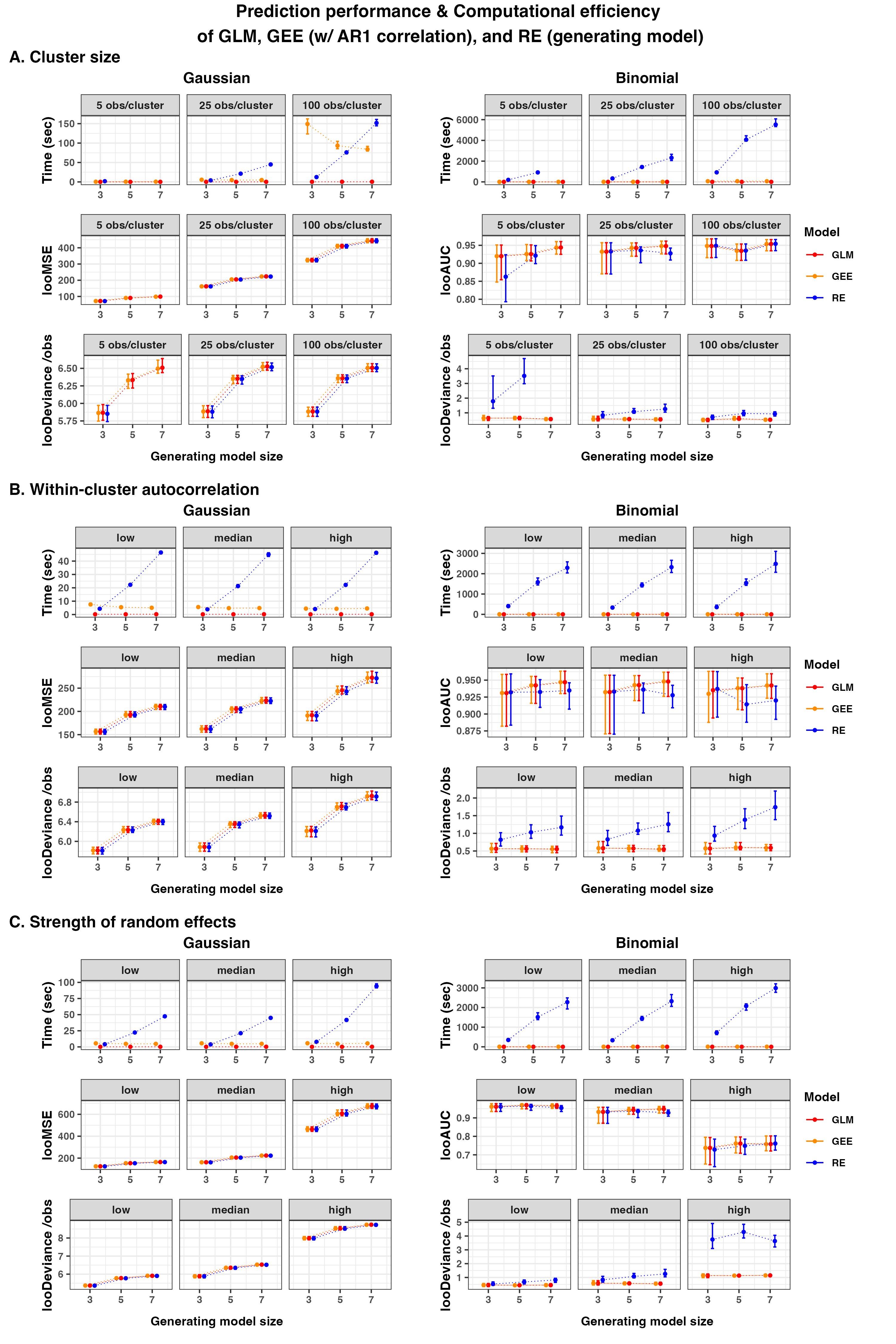}}
\caption{Leave-one-cluster-out predictive performance and computational efficiency of GLM, GEE with AR1 working correlation structure, and RE methods, for clustered data with Gaussian or binomial responses. Colors indicate different modeling methods. \textbf{Panel A} illustrates varying cluster sizes under the median levels of autocorrelation ($\phi=0.4$) and random effects ($r_b = 1$). \textbf{Panel B} illustrates varying within-cluster autocorrelation strength under the condition of 25 observations per cluster and the median level of random effects ($r_b = 1$). \textbf{Panel C} illustrates varying strength of random effects under the condition of 25 observations per cluster and the median level of autocorrelation ($\phi=0.4$). For each panel, the first row compares the time consumed in seconds to conduct a leave-one-cluster-out cross-validation by each method, the second row compares the leave-one-cluster-out mean squared error (looMSE) for Gaussian and the leave-one-cluster-out AUROC (looAUC) for binomial responses, and the third row compares the leave-one-cluster-out deviance (looDeviance), normalized by the total number of observations. The x-axis represents the total number of predictors (generating model size). In each tile, the points correspond to the median value over 100 iterations under each condition, and the error bar around each point represents the 2.5th and 97.5th percentiles.}\label{ooc_performance}
\end{figure}

\begin{figure}[htp]
    \centering
    \makebox[\textwidth]{\includegraphics[width=\textwidth]{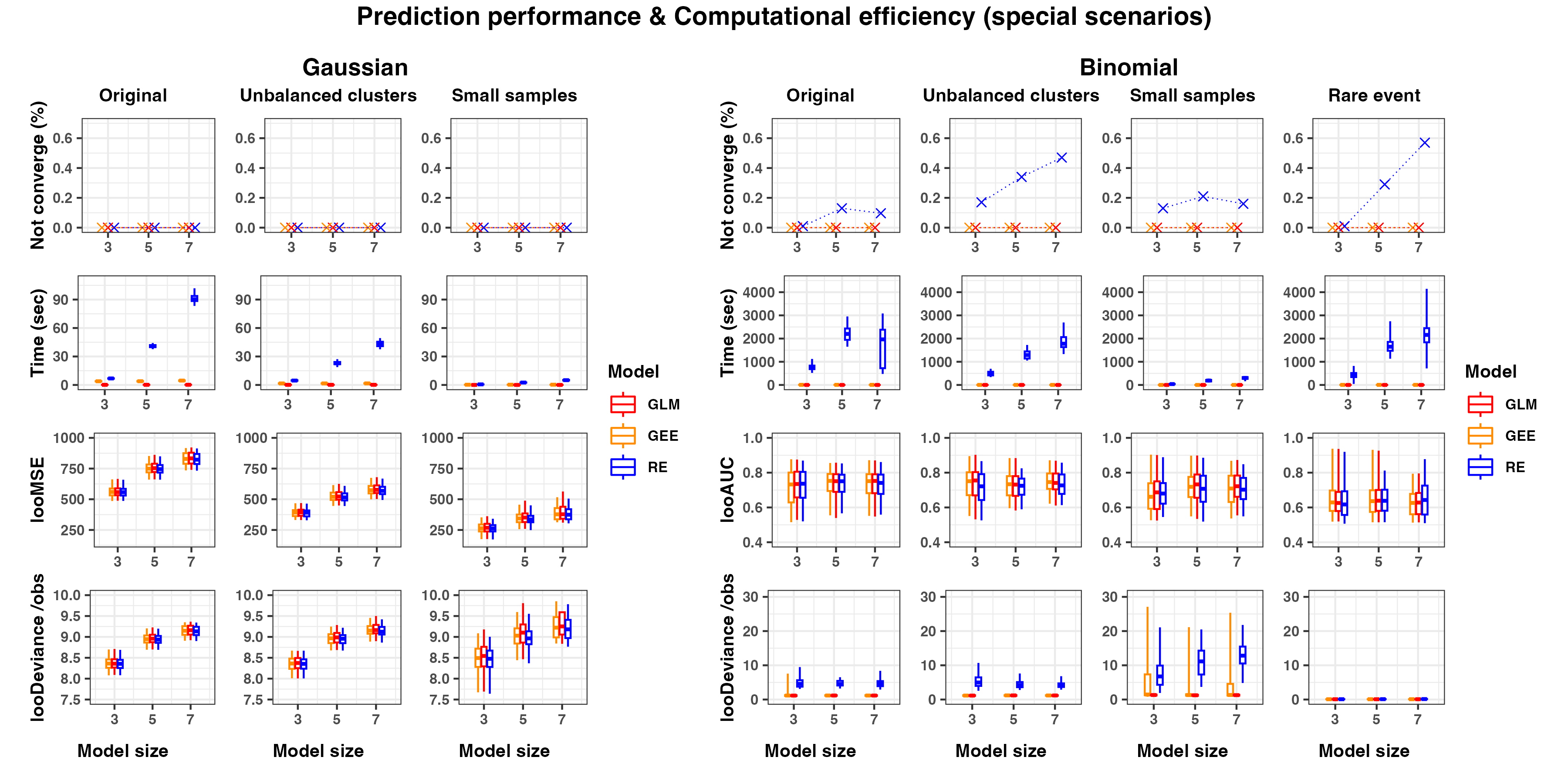}}
\caption{Leave-one-cluster-out predictive performance and computational efficiency of GLM, GEE with AR1 working correlation structure, and RE methods for clustered data with Gaussian (left panel) or binomial (right panel) responses, under special scenarios of unbalanced cluster sizes, small samples, and rare events. Clustered data were under strong clustering conditions with high autocorrelation ($\phi=0.8$), random effects ($r_b = 10$), and 50 clusters, except for the small sample scenario, which had 10 clusters. Colors indicate different modeling methods. For each panel, the first row illustrates the non-convergence rate (\%) over 100 iterations for each method under each condition. The second row compares the time consumed in seconds to conduct a leave-one-cluster-out cross-validation by each method. The third row compares the leave-one-cluster-out mean squared error (looMSE) for Gaussian and the leave-one-cluster-out AUROC (looAUC) for binomial responses. The fourth row compares the leave-one-cluster-out deviance (looDeviance), normalized by the total number of observations. The x-axis represents the total number of predictors (generating model size). In each tile, the box plot corresponds to the 2.5th, 25th, 50th, 75th, and 97.5th percentiles of measurements over 100 iterations.}\label{ooc_performance_special}
\end{figure}

\end{appendices}

\clearpage
\bibliography{ref}

\end{document}